\documentclass[11pt,a4paper]{article}

\usepackage{graphicx}
\usepackage{color}
\usepackage{amsmath,amssymb}
\usepackage{bbold}
\usepackage{t1enc}
\usepackage{cite}

\usepackage{mathtools}
\usepackage{yhmath}
\usepackage{amsmath,bm,amssymb}
\usepackage{amsfonts}
\usepackage{bbold}
\usepackage{mathrsfs}
\usepackage{latexsym} 
\usepackage{cancel}
\usepackage{bm}
\usepackage{graphicx, rotating}
\usepackage{epstopdf}
\usepackage{epsfig}
\usepackage{latexsym}
\usepackage{color}
\usepackage[dvipsnames]{xcolor}

\DeclareMathSymbol{\shortminus}{\mathbin}{AMSa}{"39}
\DeclareMathSymbol{\shm}{\mathbin}{AMSa}{"39}
\newcommand{\oh}{\frac{1}{2}}
\newcommand{\toh}{\textstyle \frac{1}{2}}

\newcommand{\UU}{|{ \small ++} \rangle}
\newcommand{\UD}{|{ \small +-} \rangle}
\newcommand{\UZ}{|{ \small +0} \rangle}
\newcommand{\DU}{|{ \small -+} \rangle}
\newcommand{\DD}{|{ \small --} \rangle}
\newcommand{\DZ}{|{ \small -0} \rangle}

\begin{document}

\begin{center}
\begin{Large}
{\bf A closer look at post-decay $t \bar t$ entanglement}
\end{Large}

\vspace{0.5cm}
\renewcommand*{\thefootnote}{\fnsymbol{footnote}}
\setcounter{footnote}{0}
J.~A.~Aguilar-Saavedra \\[1mm]
\begin{small}
Instituto de F\'isica Te\'orica IFT-UAM/CSIC, c/Nicol\'as Cabrera 13--15, 28049 Madrid, Spain \\
\end{small}
\end{center}

\begin{abstract}
Top pair production is ideally suited to observe post-decay entanglement, thus providing a novel test of quantum mechanics. We provide top polarised decay amplitudes that can be used to compute semi-analytical predictions and, in particular, to better understand {\em why} the post-decay entanglement arises. We obtain predictions for the LHC, identifying general phase space regions where experimental measurements of $tW$ entanglement are feasible. We also give predictions for polarised $e^+ e^-$ collisions, focusing on the possibility that the post-decay $tW$ entanglement is larger than the $t \bar t$ one.
\end{abstract}

\section{Introduction}

Quantum entanglement between heavy particles produced in high-energy collisions at the Large Hadron Collider (LHC) has recently become a fruitful field of research, with proposals to test the spin entanglement of top pairs~\cite{Afik:2020onf,Fabbrichesi:2021npl,Severi:2021cnj,Afik:2022kwm,Aguilar-Saavedra:2022uye,Afik:2022dgh,Dong:2023xiw,Han:2023fci}, weak boson pairs produced  in the decay of a Higgs boson~\cite{Barr:2021zcp,Aguilar-Saavedra:2022wam,Aguilar-Saavedra:2022mpg,Fabbri:2023ncz}, and via electroweak interactions~\cite{Ashby-Pickering:2022umy,Fabbrichesi:2023cev,Morales:2023gow}. Measurements at the LHC can extend the energy range in which quantum entanglement has been tested, and this has already been achieved with the first measurements of $t \bar t$ entanglement by the ATLAS and CMS Collaborations~\cite{note,note2}. But, remarkably, elementary particle physics can also provide a novel type of tests not yet performed: those that involve quantum entanglement and particle decay, that is, post-decay entanglement~\cite{Aguilar-Saavedra:2023hss}. Namely, testing how the spin entanglement between two particles $A$ and $B$ is inherited by their decay products. 

The framework to provide predictions is that of post-decay density operators~\cite{Aguilar-Saavedra:2024fig}. In this respect, we note that previous literature has often addressed predictions of decay angular distributions for heavy particle decays, using in particular the helicity formalism of Jacob and Wick~\cite{Jacob:1959at}. But, although the spin information is implicit\footnote{From the fully-differential distributions one can determine the spin state of all particles involved, by performing phase space integrations with appropriate kernels.} in angular distributions, it is not {\em explicit}. The formalism of post-decay operators focuses on direct calculations of the spin state of the decay products, which arises from total angular momentum conservation and the measurement of the momenta of decay products. This approach provides a deeper understanding of the physics involved. In particular, it has allowed to find that the entanglement between decay products can be larger than that of the parent particles~\cite{Aguilar-Saavedra:2024fig}.

The best playground to test post-decay entanglement at high-energy colliders is $t \bar t$ production. Thus, after a brief review of the formalism in section~\ref{sec:2}, we provide polarised top quark decay amplitudes in section~\ref{sec:3}. Though helicity amplitudes have previously been computed~\cite{Boudjema:2009fz,Aguilar-Saavedra:2024fig}, polarised amplitudes using a fixed reference system are necessary for general calculations of post-decay density operators. Section~\ref{sec:4} is devoted to a better understanding of the post-decay $t \bar t$ entanglement and its relation to the $W$ spin analysing power.
Predictions for $t \bar t$ production at the LHC are given in section~\ref{sec:5}, in particular identifying how the amount of entanglement depends on the phase space region considered for the top (anti-)quark decay. The entanglement at $e^+ e^-$ colliders is discussed in section~\ref{sec:6}, and our results summarised in section~\ref{sec:7}. A comparison between the results obtained using this formalism, and Monte Carlo simulations, is given in appendix~\ref{sec:a}.


\section{Post-decay density operators}
\label{sec:2}

Here we summarise the main results of the post-decay density operator formalism~\cite{Aguilar-Saavedra:2024fig}.
Let us consider a system of two particles $A$ (Alice) and $B$ (Bob). In the most general case, the spin state is described by a density operator
\begin{equation}
\rho = \sum_{ijkl} \rho_{ij}^{kl} |\phi_i \chi_k \rangle \langle \phi_j \chi_l | \,.
\end{equation}
where  $|\phi_i \rangle$ and $|\chi_k\rangle$ are bases to parameterise the spin states of $A$ and $B$, respectively. We consider $A$ decaying into some given final state, $A \to A_1 \dots A_n$, while the possible decay of $B$ is not considered. The matrix elements for this decay are
\begin{equation}
M_{\alpha j} \equiv \langle P \, \xi_\alpha | T | \phi_j \rangle \,,
\label{ec:Mij}
\end{equation}
with $\xi_\alpha$ generically denoting spin indices of the multi-particle final state, and $P$ labelling the momenta of $A_1, \dots,  A_n$. The spin state after the decay of $A$ is obtained by projecting the decayed state on the subspace of fixed momenta $|P\rangle$, and is given by
\begin{equation}
\rho' = \frac{1}{\sum_{\alpha k} (M \rho^{kk} M^\dagger)_{\alpha \alpha}} 
\sum_{\alpha \beta kl} (M \rho^{kl} M^\dagger)_{\alpha \beta} |\xi_\alpha \chi_k \rangle \langle \xi_\beta \chi_l | \,,
\label{ec:rhop}
\end{equation}
with matrix multiplication in the lower indices of $\rho_{ij}^{kl}$. In case $|P \rangle$ is integrated over some region $S$, the density operator is 
\begin{equation}
\rho' = \frac{1}{\sum_{\alpha k} \int_S d\Omega \, (M \rho^{kk} M^\dagger)_{\alpha \alpha}} 
\sum_{\alpha \beta kl} \left[ \int_S d\Omega \, (M \rho^{kl} M^\dagger)_{\alpha \beta} \right] |\xi_\alpha \chi_k \rangle \langle \xi_\beta \chi_l | \,.
\label{ec:rhopint}
\end{equation}
In our notation we have implicitly assumed that $A$ undergoes a two-body decay, in which case the phase space is two-dimensional and can be parameterised by two angles $\Omega=(\theta,\phi)$. Otherwise, the generalisation to multi-body decays is straightforward.

Specifically for the $t \bar t$ system, as basis of states $|\phi_i \chi_k \rangle$ we use the basis of $S_3$ eigenstates
\begin{equation}
\{  |\toh \, \toh \rangle \,, |\toh \, -\!\toh \rangle \,, |-\! \toh \, \toh \rangle \,, 
|-\!\toh \, -\!\toh \rangle \}
\label{ec:ttbasis}
\end{equation}
where the first particle is the quark and the second particle the anti-quark.
The spin states $|\xi_\alpha \rangle$ in this case are those of the $b W^+$ ($\bar b W^-$) pair for $t$ ($\bar t$) decays,
\begin{equation}
\{  |\toh \, 1 \rangle \,, |\toh \, 0 \rangle \,, |\toh \, -\!1 \rangle \,, 
|-\!\toh \, 1 \rangle \,, |-\!\toh \, 0 \rangle \,, |-\! \toh \, -\!1 \rangle \} \,.
\label{ec:tWbasis}
\end{equation}


\section{Top polarised decay amplitudes}
\label{sec:3}

We consider the decay of a top quark $t \to Wb$ in its rest frame, in which we take a fixed reference system $(x,y,z)$.  The three-momentum of the $W$ boson is parameterised as $\vec p = q (\sin \theta \cos \phi,\sin \theta \sin \phi,\cos \theta)$ and $E_W$, $E_b$ are the energies of $W$ and $b$, respectively, all quantities evaluated in the top quark rest frame. The masses of $t$, $W$ and $b$ are denoted as $m_t$, $M_W$ and $m_b$, as usual. Unless otherwise indicated, we take $m_t = 172.5$ GeV, $M_W = 80.4$ GeV, $m_b = 4.8$ GeV, and consider a generic interaction
\begin{equation}
\mathcal{L} = - \frac{1}{\sqrt 2} \bar b \gamma^\mu (g_L P_L + g_R P_R) t \, W_\mu^- 
= - \frac{1}{2 \sqrt 2} \bar b \gamma^\mu (g_V - g_A \gamma_5) t \, W_\mu^- \,.
\label{ec:lagr}
\end{equation}
The standard model (SM) interaction is recovered by setting $g_L = g$, $g_R = 0$ or $g_V = g_A = g$. All the spins are quantised along the $\hat z$ axis, and we label the amplitudes as $A_{s_1\,s_2\,s_3}$, where $s_1$, $s_2$, $s_3$ are the third spin components of $t$, $b$ and $W$, respectively. This contrasts with helicity amplitudes, in which the spin of the decay products is quantised in their direction of motion.
The $W$ rest-frame polarisation vectors are
\begin{align}
& \varepsilon_R^{(+)} = - \frac{1}{\sqrt 2} (0,1,i,0) \,, \notag \\
& \varepsilon_R^{(0)} = (0,0,0,1) \,, \notag \\
& \varepsilon_R^{(-)} = \frac{1}{\sqrt 2} (0,1,-i,0) \,.
\label{ec:epsR}
\end{align}
From these, the polarisation vectors in the top quark rest frame can be obtained with a boost,
\begin{align}
& \varepsilon^{(+)} = \varepsilon_R^{(+)} - \frac{1}{\sqrt 2} \sin \theta e^{i \phi} \frac{q}{M_W} \left( 1, \frac{\vec p}{M_W+E_W} \right) \,, \notag \\
& \varepsilon^{(0)} = \varepsilon_R^{(0)} + \cos \theta \frac{q}{M_W} \left( 1, \frac{\vec p}{M_W+E_W} \right) \,, \notag \\
& \varepsilon^{(-)} = \varepsilon_R^{(-)} + \frac{1}{\sqrt 2} \sin \theta e^{-i \phi} \frac{q}{M_W} \left( 1, \frac{\vec p}{M_W+E_W} \right) \,.
\end{align}
The amplitudes can easily be computed by using the explicit expression of the $t$ and $b$ spinors in the Dirac basis and the above polarisation vectors. For convenience, we introduce the kinematical factors
\begin{align}
& F = [m_t (E_b+m_b)]^{1/2} \frac{q}{E_b+m_b} \,, \notag \\
& F_0 = [m_t (E_b+m_b)]^{1/2}  \left[ 1 + \frac{E_W+M_W}{M_W} \left( 1 + \frac{E_W - M_W}{E_b+m_b} \right) \right] \,,
\notag \\
& F_1^\pm = [m_t (E_b+m_b)]^{1/2} \frac{q}{M_W} \left( 1 + \frac{E_W \pm M_W}{E_b+m_b} \right) \,, \notag \\
& F_1^0 = [m_t (E_b+m_b)]^{1/2} \frac{q}{M_W} \left( 1 + \frac{E_W}{E_b+m_b} \right) \,, \notag \\
& F_2 = [m_t (E_b+m_b)]^{1/2} \frac{E_W-M_W}{M_W} \left( 1 + \frac{E_W + M_W}{E_b+m_b} \right) \,.
\end{align}
In terms of the spherical harmonics $Y_l^m(\theta,\phi)$, the tree-level amplitudes read
\begin{align}
& A_{\oh\,\oh\,1} = - \sqrt{\frac{\pi}{3}} g_V F_1^+ Y_1^{-1} -  \sqrt{\frac{\pi}{15}} g_A F_2 Y_2^{-1} \,, \notag \\
& A_{\oh\,\oh\,0} = \sqrt{ \frac{\pi}{9}} g_A F_0 Y_0^0 + \sqrt{\frac{ \pi}{3}} g_V F_1^0 Y_1^0 + \sqrt{\frac{4\pi}{45}} g_A F_2 Y_2^0 \,, \notag \\
& A_{\oh\,\oh\,-\!1} = - \sqrt{\frac{\pi}{3}} g_V F_1^- Y_1^1 -  \sqrt{\frac{\pi}{15}} g_A F_2 Y_2^1 \,, \notag \\
& A_{\oh\,-\!\oh\,1} = - \sqrt{\frac{2\pi}{9}} g_A F_0 Y_0^0 + \sqrt{\frac{2\pi}{3}} g_V F Y_1^0 + \sqrt{\frac{2\pi}{45}} g_A F_2 Y_2^0 \,, \notag \\
& A_{\oh\,-\!\oh\,0} = - \sqrt{\frac{2\pi}{3}} g_V F Y_1^1 - \sqrt{\frac{2\pi}{15}} g_A F_2 Y_2^1 \,, \notag \\
& A_{\oh\,-\!\oh\, -\!1} = \sqrt{\frac{4\pi}{15}} g_A F_2 Y_2^2 \,, \notag \\
& A_{-\!\oh\,\oh\,1} = - \sqrt{\frac{4\pi}{15}} g_A F_2 Y_2^{-2} \,, \notag \\
& A_{-\!\oh\,\oh\,0} = - \sqrt{\frac{2\pi}{3}} g_V F Y_1^{-1} + \sqrt{\frac{2\pi}{15}} g_A F_2 Y_2^{-1} \,, \notag \\
& A_{-\!\oh\,\oh\,-\!1} = \sqrt{\frac{2\pi}{9}} g_A F_0 Y_0^0 + \sqrt{\frac{2\pi}{3}} g_V F Y_1^0 - \sqrt{\frac{2\pi}{45}} g_A F_2 Y_2^0 \,, \notag \\
& A_{-\!\oh\, -\!\oh\,1} = - \sqrt{\frac{\pi}{3}} g_V F_1^- Y_1^{-1} + \sqrt{\frac{\pi}{15}} g_A F_2 Y_2^{-1} \,, \notag \\
& A_{-\oh\,-\!\oh\,0} = - \sqrt{\frac{\pi}{9}} g_A F_0 Y_0^0 + \sqrt{\frac{\pi}{3}} g_V F_1^0 Y_1^0 - \sqrt{\frac{4\pi}{45}} g_A F_2 Y_2^0 \,, \notag \\
& A_{-\oh\,-\!\oh\,-1} = - \sqrt{\frac{\pi}{3}} g_V F_1^+ Y_1^1 + \sqrt{\frac{\pi}{15}} g_A F_2 Y_2^1 \,.
\label{ec:amp}
\end{align}
Angular momentum conservation is apparent here, by noting the $m$ components of the spherical harmonics that contribute to each amplitude. Note also that the highest-order spherical harmonics have $l=2$, also in agreement with angular momentum conservation. The parity-odd spherical harmonics $Y_1^m$ always have the coefficient $g_V$, while the parity-even ones $Y_0^0$, $Y_{2}^m$ have a coefficient $g_A$. In contrast to helicity amplitudes~\cite{Boudjema:2009fz,Aguilar-Saavedra:2024fig} none of the polarised amplitudes is identically zero, because of the non-trivial role that orbital angular momentum plays for a fixed basis.

The amplitudes $\bar A$ for the decay of top anti-quarks can be obtained likewise. We have calculated them explicitly as a cross-check, and find that they relate to the former by
\begin{equation}
\bar A_{s_1\,s_2\,s_3}(\theta,\phi) = -A_{s_1\,s_2\,s_3}(\pi-\theta,\phi+\pi) \,.
\end{equation}
This is expected, up to the global minus sign, from CP conservation of the tree-level $tbW$ interaction. Notice that $Y_l^m(\theta,\phi) = (-1)^l Y_l^m(\pi-\theta,\phi+\pi)$, therefore the amplitudes for $\bar t \to W \bar b$ read as (\ref{ec:amp}) but with the sign of the $Y_0^0$ and $Y_2^m$ terms changed.


\section{Understanding $t \bar t$ post-decay entanglement}
\label{sec:4}

We use the polarised amplitudes in section~\ref{sec:3} and the framework of post-decay density operators to understand how the $t \bar t$ entanglement is inherited by their decay products. For definiteness we will consider the decay of the anti-quark, namely, $tW^-$ entanglement. 
The decay amplitudes $M_{\alpha j}$ defined in (\ref{ec:Mij}) can be read from $A_{s_1\,s_2\,s_3}$ in Eqs.~(\ref{ec:amp}) with the appropriate replacements for $\bar t$ decays, identifying $\alpha=1,\dots,6$ with the spin indices $(s_2,s_3)$ for $\bar b W^-$, ordered as in (\ref{ec:tWbasis}) and $j=1,2$ with the $\bar t$ spin index $s_1$. Unless otherwise noted, we assume the SM value of the coupling.
Taking the partial trace in the $\bar b$ space, the $t W^-$ density operator is obtained.

The entanglement is quantified by two measures~\cite{Plenio:2007zz}. The first one is the negativity of the partial transpose on the $B$ subspace $\rho^{T_B}$,
\begin{equation}
N(\rho) = \frac{\| \rho^{T_B} \| - 1}{2} \,,
\label{ec:Nrho}
\end{equation}
where $\|X\| = \operatorname{tr} \sqrt{X X^\dagger} = \sum_i \sqrt{\lambda_i}$, where $\lambda_i$ are the (positive) eigenvalues of the matrix $X X^\dagger$. Because $\rho^{T_B}$ has unit trace, $N(\rho)$ equals the sum of the negative eigenvalues of $\rho^{T_B}$. The second measure is the logarithmic negativity
\begin{equation}
E_N(\rho) = \log_2 \| \rho^{T_B} \| \,.
\end{equation}
In either case, the measures can also be computed by taking the partial transpose in the $A$ subspace, obtaining the same result.

Let us consider the $t \bar t$ density operator corresponding to the state $|\psi\rangle = \left[ |\toh \, \toh \rangle + |-\! \toh \, -\!\toh \rangle \right]$, given by
\begin{equation}
\rho = \frac{1}{2} \left( \! \begin{array}{cccc}
1 & 0 & 0 & 1 \\
0 & 0 & 0 & 0 \\
0 & 0 & 0 & 0 \\
1 & 0 & 0 & 1
\end{array} \! \right) \,.
\label{ec:rhottex}
\end{equation}
This operator has $N = 0.5$, $E_N = 1$. The $tW^-$ density operator depends on $(\theta,\phi)$ via the amplitudes $M_{\alpha j}$ We illustrate the post-decay $tW^-$ entanglement selecting $\vec p$ along any of the positive or negative axes. In those cases,
\begin{align}
& \vec p \, \| \pm \hat z \,,~  \vec p \, \| \pm \hat x \,,~
\vec p \, \| \pm \hat y
&& N = 0.459 \,, E_N = 0.940  \,.
\label{ec:med80}
\end{align}
We observe that the $tW^-$ entanglement is quite close to the $t \bar t$ one. For the specific case of top quark decays, this feature can be related to the spin analysing power of the $W$ boson.\footnote{The spin analysing power $\kappa_W$ is a constant quantifying the dependence of the $W$ angular distribution on the top quark polarisation, e.g. $(1/\Gamma) d\Gamma/d\cos \theta = 1/2(1+ \alpha_W P_3 \cos \theta)$, with $P_3 = 2 \langle S_3 \rangle$~\cite{Jezabek:1994zv}.} 
Its relatively small value, $\kappa_W = 0.39$ at the tree level, is due to the fact that there is not a significant dominance of amplitudes with $\bar t$ spin $1/2$ or $-1/2$. For example, for $\theta = 0,\pi$ the amplitudes $M_{\alpha j}$ are
\begin{align}
& M^T = \left( \! \begin{array}{cccccc}
0 & -1.15 & 0 & 98.3 & 0 & 0 \\
0 & 0 & -3.5 & 0 & 149 & 0
\end{array} \! \right) \quad (\theta = 0) \,, \notag \\[2mm]
& M^T = \left( \! \begin{array}{cccccc}
0 & -149 & 0 & 3.5 & 0 & 0 \\
0 & 0 & -98.3 & 0 & 1.15 & 0
\end{array} \! \right)  \quad (\theta = \pi) \,, \label{ec:MTpi}
\end{align}
in units of GeV. Notice the suppression of $\bar b$ with helicity $-1/2$, namely the first three columns of $M^T$ for $\theta = 0$ and the last three columns for $\theta = \pi$. The fact that the momentum measurement does not produce a clear preference for either $\bar t$ spin is translated into a significant $tW^-$ entanglement. 
In this respect, note that for $\bar b$ with positive helicity there are only two possible spin states for the $W$ boson, depending on the spin of the parent $\bar t$ quark, e.g.
\begin{align}
& \bar t~ |\toh \rangle  \to \bar bW^-\; |-\!\toh \, 1 \rangle \,; \quad
\bar t~ |-\! \toh \rangle \to \bar bW^-\; |-\! \toh \, 0 \rangle && (\theta = 0) \,, \notag \\
& \bar t~ |\toh \rangle \to \bar bW^-\; | \toh \, 0 \rangle \,; \quad
\bar t~ |-\! \toh \rangle \to \bar bW^-\;  | \toh \, -\!1 \rangle && (\theta = \pi) \,.
\end{align}
Therefore, an entangled $t \bar t$ state that has components with $\bar t$ of either spin produces, after decay, an entangled state where $bW$ have the spin combinations shown above, and the fact that neither amplitudes dominate maintains the entanglement to a large extent. Besides, note also that the symmetry of the initial state results in identical entanglement measures for $\vec p$ along the $\pm \hat z$, $\pm \hat x$ and $\pm \hat y$ axes.

The relation between entanglement and spin analysing power can be further clarified by taking decay amplitudes with other parameters. If we set $M_W = 122$ GeV, the $W$ boson spin analysing power nearly vanishes, $\kappa_W = -6 \times 10^{-4}$. The decay amplitudes are
\begin{align}
& M^T = \left( \! \begin{array}{cccccc}
0 & -2.19 & 0 & 78.5 & 0 & 0 \\
0 & 0 & -4.38 & 0 & 78.3 & 0
\end{array} \! \right) \quad (\theta = 0) \,, \notag \\[2mm]
& M^T = \left( \! \begin{array}{cccccc}
0 & -78.3 & 0 & 4.37 & 0 & 0 \\
0 & 0 & -78.5 & 0 & 2.19 & 0
\end{array} \! \right)  \quad (\theta = \pi) \,
\end{align}
in units of GeV. There is equal balance for amplitudes with $\bar t$ spin $1/2$ and $-1/2$, and the $tW^-$ entanglement is practically the same as the $t \bar t$ one,
\begin{align}
& \vec p \, \| \pm \hat z \,,~  \vec p \, \| \pm \hat x \,,~
\vec p \, \| \pm \hat y
&& N = 0.499 \,, E_N = 0.998  \,.
\end{align}
On the other hand, if we set $M_W = 5$ GeV, the spin analysing power is nearly maximal, $\kappa_W = 0.997$. The amplitudes (in the same units of GeV) are in this case
\begin{align}
& M^T = \left( \! \begin{array}{cccccc}
0 & -0.06 & 0 & 111 & 0 & 0 \\
0 & 0 & -3.09 & 0 & 2707 & 0
\end{array} \! \right) \quad (\theta = 0) \,, \notag \\[2mm]
& M^T = \left( \! \begin{array}{cccccc}
0 & -2707 & 0 & 3.09 & 0 & 0 \\
0 & 0 & -111 & 0 & 0.06 & 0
\end{array} \! \right)  \quad (\theta = \pi) \,.
\end{align}
We observe a clear dominance for $\bar t$ spin $S_3 = -1/2$ for $\theta = 0$, and $S_3 = 1/2$ for $\theta = \pi$. As a result, the $tW^-$ entanglement nearly vanishes,
\begin{align}
& \vec p \, \| \pm \hat z \,,~  \vec p \, \| \pm \hat x \,,~
\vec p \, \| \pm \hat y
&& N = 0.040 \,, E_N = 0.113  \,.
\end{align}

\begin{figure}[htb]
\begin{center}
\includegraphics[height=5.5cm,clip=]{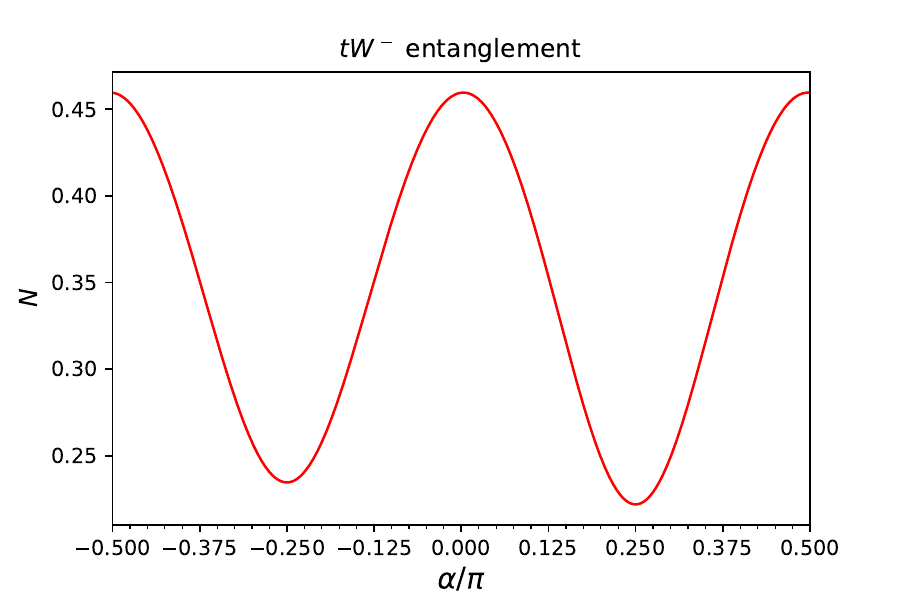} 
\caption{Entanglement measure $N$ as a function of $\alpha$ in Eq.~(\ref{ec:galpha}).}
\label{fig:Nvscoup}
\end{center}
\end{figure}

The amplitudes given in section~\ref{sec:3} also allow to analyse in detail the impact of the $tbW$ coupling chirality on the $tW^-$ entanglement.
We parameterise the couplings appearing in Eq.~(\ref{ec:lagr}) as
\begin{equation}
g_L = g \cos \alpha \,,~ g_R = g \sin \alpha \,.
\label{ec:galpha}
\end{equation}
In Fig. \ref{fig:Nvscoup} we plot $N$ for $\vec p \, \| \hat z$ ($\theta = 0$) as a function of $\alpha$, i.e. fixing the remaining parameters. The SM corresponds to $\alpha = 0$, where $N$ reaches the global maximum of $N(\alpha)$.  With a right-handed coupling, $\alpha = \pm \pi/2$, $N$ takes the same value. The entanglement is minimal for $\alpha = \pi/4$, corresponding to a vector coupling. An axial coupling $\alpha = - \pi/4$ is a local minimum of $N(\alpha)$, with practically the same value as for a vector coupling. In both the vector and axial cases, the suppression of the entanglement is due to the trace over $\bar b$ spin degrees of freedom.

\section{LHC predictions for $tW$ entanglement}
\label{sec:5}

The feasibility of $tW^-$ entanglement measurements (or, equivalently, $\bar tW^+$) at the LHC has been investigated in Ref.~\cite{Aguilar-Saavedra:2023hss}. 
Here we use our framework to provide semi-analytical predictions, where the $t \bar t$ spin density operator is obtained from a Monte Carlo calculation and the polarised decay amplitudes are those in section~\ref{sec:3}. We give examples for the two kinematical regions (near threshold and boosted central tops) previously studied in Refs.~\cite{Aguilar-Saavedra:2022uye,Aguilar-Saavedra:2023hss}, and various phase space regions for the decay of the top anti-quark.

\subsection{Entanglement near threshold}
\label{sec:5.1}

Near threshold we consider $t \bar t$ pairs with invariant mass $m_{t \bar t} \leq 390$ GeV and $t \bar t$ and velocity $\beta \leq 0.9$, with 
\begin{equation}
\beta = \frac{|p_t^z + p_{\bar t}^z|}{|E_t + E_{\bar t}|}
\end{equation}
measured in the laboratory frame. Details about the simulation can be found in Ref.~\cite{Aguilar-Saavedra:2023hss}. We use the beamline basis, with $\hat x = (1,0,0)$, $\hat y = (0,1,0)$, $\hat z = (0,0,1)$. In the basis of $S_3$ given in (\ref{ec:ttbasis}), the density operator is
\begin{equation}
\rho = \left( \! \begin{array}{cccc}
0.156 & 0 & 0 & 0 \\
0 & 0.344 & -0.304 & 0 \\
0 & -0.304 & 0.344 & 0 \\
0 & 0 & 0 & 0.156
\end{array} \! \right) \,.
\label{ec:rhott}
\end{equation}
It is dominated by the spin-singlet component $1/\sqrt{2} \left[ |\toh \, -\!\toh \rangle - |-\! \toh \, \toh \rangle \right]$, with eigenvalue 0.65. The entanglement measures are $N = 0.15$, $E_N = 0.37$.

We select several kinematical regions (in terms of $\vec p$), e.g. with $\vec p$ along the positive or negative $\hat z$ axis, with $\vec p$ above or below the $xy$ plane, etc. For each region, we collect in table~\ref{tab:ent1} the principal eigenvector and largest eigenvalue $\lambda_\text{max}$ of the $tW^-$ density spin operator, as well as the entanglement measures. We use the basis of $S_3$ eigenvectors as usual, denoting the eigenstates as $|s_1 s_2\rangle$ where $s_1$ and $s_2$ are the eigenvalues for the top and $W^-$ boson, respectively.

\begin{table}[htb]
\begin{center}
\begin{tabular}{llccc}
& Principal eigenvector & $\lambda_\text{max}$ & $N$ & $E_N$ \\
$\vec p \, \| \hat z$ & $0.85 \UZ - 0.53 \DU$ & 0.65 & 0.13 & 0.34 \\
$\vec p \, \| -\hat z$ & $0.53 \UD - 0.85 \DZ$ & 0.65 & 0.13 & 0.34 \\
$p_3 > 0$ & $0.67 \UZ - 0.75 \DU$ & 0.44 & 0.07 & 0.19 \\
$p_3 < 0$ & $0.75 \UD - 0.67 \DZ$ & 0.44 & 0.07 & 0.19 \\
$\vec p \, \| \hat x$ & $0.61 [\DU - \UD] + 0.28 [\UU - \DD] + 0.23 [\DZ - \UZ]$ & 0.68 & 0.13 & 0.34 \\
$\vec p \, \| -\hat x$ & $0.61 [\DU + \UD] - 0.28 [\UU + \DD] - 0.23 [\DZ + \UZ]$ & 0.68 & 0.13 & 0.34 \\
$p_1 > 0$ & $0.59 [\DU - \UD] + 0.07 [\UU - \DD] + 0.38 [\DZ - \UZ]$ & 0.44 & 0.07 & 0.18 \\
$p_1 < 0$ & $0.59 [\DU + \UD] - 0.07 [\UU + \DD] - 0.38 [\DZ + \UZ]$ & 0.44 & 0.07 & 0.18 \\
$\vec p \, \| \hat y$ & $0.61 [\DU - i \UD] + 0.28 [\DD - i \UU] - 0.23 [\UZ - i \DZ]$ & 0.68 & 0.13 & 0.34 \\
$\vec p \, \| - \hat y$ & $0.61 [\DU + i \UD] + 0.28 [\DD + i \UU] - 0.23 [\UZ + i \DZ]$ & 0.68 & 0.13 & 0.34 \\
$p_2 > 0$ & $0.59 [\DU - i \UD] + 0.07 [\DD - i \UU] - 0.38 [\UZ - i \DZ]$ & 0.44 & 0.07 & 0.18 \\
$p_2 < 0$ & $0.59 [\DU + i \UD] + 0.07 [\DD + i \UU] - 0.38 [\UZ + i \DZ]$ & 0.44 & 0.07 & 0.18 \\
\end{tabular}
\caption{Principal eigenvector and largest eigenvalue of the $tW^-$ spin density operator, and entanglement measure, for the near-threshold region at LHC. We write in compact notation the $1/2$ and $1$ spin eigenvalues of $t$ and $W^-$ as `$+$', and the $-1/2$, $-1$ eigenvalues as `$-$'.} 
\label{tab:ent1}
\end{center}
\end{table}

From these examples we observe that $tW^-$ are equally entangled when $\vec p$ is in either of the axes, not only along the positive or negative $\hat z$ axis. (Actually, the value of $N$, $E_N$ are the same for every direction of $\vec p$.) When $\vec p \, \| \hat z$ or $\vec p \, \| -\hat z$, the principal eigenvectors have only two components in the basis of $S_z$ eigenvectors, as expected from angular momentum conservation~\cite{Aguilar-Saavedra:2023hss}. It is perhaps more surprising to observe that the principal eigenvectors still have the same two components in the half-spaces $p_3 > 0$ and $p_3 < 0$, respectively. This is due to the particular form of the $t \bar t$ density operator (\ref{ec:rhott}) and the integration over $\phi$. In the remaining cases, the principal eigenvectors have non-zero components in the six basis vectors. However, by a change of basis their expressions are simplified. For example, for $\vec p \, \| \hat x$ the principal eigenvector is $0.89 \UZ_x - 0.46 \DU_x$, with the $x$ subindex referring to the basis of $S_1$. The similarity with the case $\vec p \, \| \hat z$ is of course due to the fact that the $t \bar t$ density operator $\rho$ is dominated by the spin-singlet component.

Another remarkable point is that the entanglement decreases when $\vec p$ is integrated over some region. This may seem paradoxical, because the entanglement is the same for any direction of $\vec p$. However, the actual quantum states are different and integrating over a region as in (\ref{ec:rhopint}) reduces the entanglement. When integrating over the full phase space for $\bar t$ decay, it is found that $t$ and $W^-$ are no longer entangled. In this case there are three dimension-2 degenerate subspaces. The one with the largest eigenvalue (0.31) is spanned by the two vectors $0.63 \UZ -0.78 \DU$ and $0.78 \UD - 0.63 \DZ$. For an experimental measurement $\vec p$ has to be considered in some relatively narrow region, e.g. a cone around some value $\vec p_0$, in order to have sufficient statistics. It is found that for fixed cone radius the entanglement is the same for any orientation of this cone (that is, of $\vec p_0$). This property may constitute an advantage to avoid possible reconstruction and efficiency issues that might be present in some specific phase space region. Moreover, measurements in different non-overlapping cones (all with the same radius) can easily be combined, in order to gain statistics. One could for example consider momenta $\vec p$ such that $\cos \angle(\vec p,\vec p_0) \geq 0.8$, for different values of $\vec p_0$. The $tW^-$ entanglement could be separately measured within these different regions and, since the theoretical prediction is the same for all values of $\vec p_0$, the results of those measurements could be statistically combined, as different measurements of the same quantity.

A final remark is in order. In order to determine the expected sensitivity to entanglement, Ref.~\cite{Aguilar-Saavedra:2023hss} used the Peres-Horodecki criterion~\cite{Peres:1996dw,Horodecki:1997vt}, selecting the lowest eigenvalue of $\rho^{T_B}$ (labelled as $\lambda_1$) as entanglement indicator. In most cases there is only one negative eigenvalue, so $N(\rho)$ shown in table~\ref{tab:ent1} equals $-\lambda_1$.

\subsection{Boosted region}
\label{sec:5.2}

Here we consider $t \bar t$ pairs with invariant mass $m_{t \bar t} \geq 800$ GeV and production angle $\theta_t$ in the range $|\cos \theta_t| \leq 0.6$. We use a `rotating' reference system with axes
$\hat z = \hat k$, $\hat x = \hat r$, $\hat y = \hat n$, the K, R and N axes being defined as
\begin{itemize}
\item K-axis (helicity): $\hat k$ is a normalised vector in the direction of the top quark three-momentum in the $t \bar t$ rest frame.
\item R-axis: $\hat r$ is in the production plane and defined as $\hat r = (\hat p_p - \cos \theta_t \hat k)/\sin \theta_t$.
\item N-axis: $\hat n = \hat k \times \hat r$ is orthogonal to the production plane.
\end{itemize}
This reference system is usually referred to as the `helicity basis'~\cite{Bernreuther:2015yna}. We use the same axes for the top quark and anti-quark. The density operator for the $t \bar t$ pair is
\begin{equation}
\rho = \left( \! \begin{array}{cccc}
0.382 & 0 & 0 & 0.288 \\
0 & 0.118 & 0.032 & 0 \\
0 & 0.032 & 0.118 & 0 \\
0.288 & 0 & 0 & 0.382
\end{array} \! \right) \,,
\label{ec:rhott2}
\end{equation}
and its principal eigenvector is $1/\sqrt{2} \left[ |\toh \, \toh \rangle + |-\! \toh \, -\!\toh \rangle \right]$, with eigenvalue 0.67. The entanglement measures are $N = 0.17$, $E_N = 0.42$.

We follow the same procedure of the previous subsection to obtain the post-decay density operator. The results for several kinematical regions (in terms of $\vec p$) are presented in table~\ref{tab:ent2}. The entanglement shows features that are similar to the near-threshold region. Note however that the entanglement for $\vec p$ along the different axes has a slight variation, smaller than the precision of the numbers given in table~\ref{tab:ent2}. This is expected because the $t \bar t$ state is not as symmetric as in the previous case, in which it is dominated by the spin-singlet component. We have also verified that for arbitrary directions of $\vec p$ the amount of entanglement is nearly the same. Therefore, an experimental measurement can be performed for $\vec p$ in a relatively narrow cone centered on any arbitrary $\vec p_0$. Measurements performed on several non-overlapping cones can also be combined in order to gain statistics. As an example, we have verified that the entanglement measures are the same for several random choices of $\vec p_0$, using cones of $\cos \angle(\vec p,\vec p_0) \geq 0.8$. 

\begin{table}[t]
\begin{center}
\begin{tabular}{llccc}
& Principal eigenvector & $\lambda_\text{max}$ & $N$ & $E_N$ \\
$\vec p \, \| \hat z$ & $0.50 \UU + 0.86 \DZ$ & 0.69 & 0.15 & 0.38 \\
$\vec p \, \| -\hat z$ & $0.86 \UZ + 0.50 \DD$ & 0.69 & 0.15 & 0.38 \\
$p_3 > 0$ & $0.75 \UU + 0.67 \DZ$ & 0.45 & 0.08 & 0.22 \\
$p_3 < 0$ & $0.67 \UZ + 0.75 \DD$ & 0.45 & 0.08 & 0.22 \\
$\vec p \, \| \hat x$ & $0.28 [\UD + \DU] - 0.61 [\UU + \DD] - 0.23 [\DZ + \UZ]$ & 0.70 & 0.15 & 0.38 \\
$\vec p \, \| -\hat x$ & $0.28 [\UD - \DU] - 0.61 [\UU - \DD] - 0.23 [\DZ - \UZ]$ & 0.70 & 0.15 & 0.38  \\
$p_1 > 0$ & $0.06 [\UD + \DU] - 0.60 [\UU + \DD] - 0.38 [\DZ + \UZ]$ & 0.45 & 0.08 & 0.22 \\
$p_1 < 0$ & $0.06 [\UD - \DU] - 0.60 [\UU - \DD] - 0.38 [\DZ - \UZ]$ & 0.45 & 0.08 & 0.22 \\
$\vec p \, \| \hat y$ & $0.25 [\DU - i \UD] + 0.61 [\DD - i \UU] + 0.25 [\UZ - i \DZ]$ & 0.68 & 0.15 & 0.38 \\
$\vec p \, \| - \hat y$ & $0.25 [\DU + i \UD] + 0.61 [\DD + i \UU] + 0.25 [\UZ + i \DZ]$ & 0.68 & 0.15 & 0.38 \\
$p_2 > 0$ & $0.07 [\DU - i \UD] + 0.59 [\DD - i \UU] + 0.38 [\UZ - i \DZ]$ & 0.45 & 0.08 & 0.22 \\
$p_2 < 0$ & $0.07 [\DU + i \UD] + 0.59 [\DD + i \UU] + 0.38 [\UZ + i \DZ]$ & 0.45 & 0.08 & 0.22 \\
\end{tabular}
\caption{Principal eigenvector and largest eigenvalue of the $tW^-$ spin density operator, and entanglement measure, for the boosted region at LHC. We write in compact notation the $1/2$ and $1$ spin eigenvalues of $t$ and $W^-$ as `$+$', and the $-1/2$, $-1$ eigenvalues as `$-$'.} 
\label{tab:ent2}
\end{center}
\end{table}


\section{Entanglement in $e^+ e^-$ collisions}
\label{sec:6}

We consider fully-polarised $e^+ e^-$ collisions at centre-of-mass energies of 0.5 TeV and 1 TeV. Top pairs are produced in an almost pure state in polarised collisions. We list in table~\ref{tab:ent4} the principal eigenvector and largest eigenvalue of the $t\bar t$ density spin operator, as well as the entanglement measures, for inclusive production and with a cut $|\cos \theta_t| \leq 0.2$, which leads to a practically pure state.

\begin{table}[htb]
\begin{center}
\begin{tabular}{lclccc}
& $\theta_t$ cut & Principal eigenvector & $\lambda_\text{max}$ & $N$ & $E_N$ \\
$e_R^+ e_L^-$ 0.5 TeV & no  & $0.27 \UU + 0.29 [\UD + \DU] + 0.87 \DD$ & 0.86 & 0.072 & 0.19 \\
$e_R^+ e_L^-$ 0.5 TeV & yes &$0.35 \UU + 0.38 [\UD + \DU] + 0.76 \DD$ & $\simeq 1$ & 0.121 & 0.31 \\
$e_R^+ e_L^-$ 1 TeV   & no  & $0.20 \UU + 0.14 [\UD + \DU] + 0.96 \DD$ & 0.91 & 0.133 & 0.34 \\
$e_R^+ e_L^-$ 1 TeV   & yes & $0.32 \UU + 0.21 [\UD + \DU] + 0.90 \DD$ & $\simeq 1$ & 0.248 & 0.58 \\
$e_L^+ e_R^-$ 0.5 TeV & no  & $0.90 \UU - 0.28 [\UD + \DU] + 0.21 \DD$ & 0.90 & 0.056 & 0.15 \\
$e_L^+ e_R^-$ 0.5 TeV & yes  & $0.80 \UU - 0.37 [\UD + \DU] + 0.30 \DD$ & $\simeq 1$ & 0.092 & 0.24 \\
$e_L^+ e_R^-$ 1 TeV   & no  & $0.97 \UU - 0.13 [\UD + \DU] + 0.15 \DD$ & 0.94 & 0.102 & 0.27 \\
$e_L^+ e_R^-$ 1 TeV   & yes  & $0.92 \UU - 0.20 [\UD + \DU] + 0.25 \DD$ & $\simeq 1$ & 0.188 & 0.46 \\
\end{tabular}
\caption{Principal eigenvector and largest eigenvalue of the $t \bar t$ spin density operator, and entanglement measure, for polarised $e^+ e^-$ collisions. We write in compact notation the $1/2$ and $-1/2$ spin eigenvalues as `$+$' and `$-$', respectively.} 
\label{tab:ent4}
\end{center}
\end{table}

\begin{table}[htb]
\begin{center}
\begin{tabular}{lcccccc}
& $\sigma$ (fb) & \multicolumn{3}{c}{$N(\rho)$} & eff. \\
&  & $t \bar t$ & $tW^-|_{\theta = \pi}$ & $tW^-|_{\cos \theta \leq -0.9}$ \\
$e_R^+ e_L^-$ 0.5 TeV inclusive                & 69.5 & 0.072 & 0.083 & 0.078 & 0.040 \\
$e_R^+ e_L^-$ 0.5 TeV $\cos \theta_t \leq 0.2$ & 12.3 & 0.121 & 0.135 & 0.129 & 0.041 \\
$e_R^+ e_L^-$ 1 TeV inclusive                  & 20.8 & 0.133 & 0.170 & 0.162 & 0.036 \\
$e_R^+ e_L^-$ 1 TeV $\cos \theta_t \leq 0.2$   & 3.3  & 0.248 & 0.313 & 0.298 & 0.037
\end{tabular}
\caption{$t \bar t$ cross section and $t \bar t$ and $tW^-$ negativity in $e_R^+ e_L^-$ collisions (see the text for details). The last column is the efficiency of the cut $\cos \theta \leq -0.9$.} 
\label{tab:EI1}
\end{center}
\end{table}

\begin{table}[t]
\begin{center}
\begin{tabular}{lcccccc}
& $\sigma$ (fb) & \multicolumn{3}{c}{$N(\rho)$} & eff. \\
&  & $t \bar t$ & $tW^-|_{\theta = 0}$ & $tW^-|_{\cos \theta \geq 0.9}$  \\
$e_L^+ e_R^-$ 0.5 TeV inclusive                & 29.6 & 0.056 & 0.068 & 0.063 & 0.038 \\
$e_L^+ e_R^-$ 0.5 TeV $\cos \theta_t \leq 0.2$ & 5.2  & 0.092 & 0.107 & 0.102 & 0.040 \\
$e_L^+ e_R^-$ 1 TeV inclusive                  & 9.3  & 0.102 & 0.139 & 0.131 & 0.034 \\
$e_L^+ e_R^-$ 1 TeV $\cos \theta_t \leq 0.2$   & 1.5  & 0.188 & 0.251 & 0.238 & 0.035
\end{tabular}
\caption{$t \bar t$ cross section and $t \bar t$ and $tW^-$ negativity in $e_L^+ e_R^-$ collisions (see the text for details). The last column is the efficiency of the cut $\cos \theta \geq 0.9$.} 
\label{tab:EI2}
\end{center}
\end{table}
As shown in Ref.~\cite{Aguilar-Saavedra:2024fig}, the $tW^-$ entanglement depends both on $\theta$ and $\phi$, and it is possible to have an entanglement amplification by selecting $\vec p$ in a suitable phase space region. In general, the normalisation of the negativity $N$ is not the same for a qubit-qubit and qubit-qutrit system. However, as previously discussed, in the limit of massless $b$ quarks the $W$ boson only has two possible spin states, and the comparison between negativities for $t \bar t$ and $tW^-$ is meaningful. Moreover, for pure states the concurrence
\begin{equation}
C^2 = 2 \left( 1 -  \operatorname{tr} \rho_A^2 \right) \,,
\end{equation}
with $\rho_A$ the reduced operator obtained after trace in the $B$ subspace,
 can be used as a measure of entanglement. In this case the dimensionality of $A$ is the same for the qubit-qubit and qubit-qutrit systems, and using this entanglement measure one also observes an entanglement amplification~\cite{Aguilar-Saavedra:2024fig}.

In $e_R^+ e_L^-$ collisions the region with maximum entanglement is typically near $\theta = \pi$, and for $e_L^+ e_R^-$ collisions near $\theta = 0$. Here we give examples for simple kinematical regions in terms of $\theta$, listing in tables~\ref{tab:EI1} and \ref{tab:EI2} the results for the negativity of the $tW^-$ pair, $N(\rho_{tW^-})$. 
For illustration we also give the $t \bar t$ cross section in the dilepton final state (considering electrons and muons) and, in the last column, the efficency of the cut $\cos \theta \leq -0.9$ or $\cos \theta \geq 0.9$, which is nearly the same for all cases. For better readability we also include the negativity of the $t \bar t$ pair, $N(\rho_{t \bar t})$.

Even though the regions $\cos \theta \leq -0.9$ or $\cos \theta \geq 0.9$ considered are not optimised (the value of $N$ also depends on $\phi$), and the cut $|\cos \theta_t| \leq 0.2$ is illustrative, it is interesting to obtain an estimate of the statistical uncertainty on $N(\rho_{tW^-})$. We concentrate on $e_R^+ e_L^-$ collisions at 1 TeV, with $\cos \theta \leq -0.9$. We collect in table~\ref{tab:stat} the statistical uncertainty of $N(\rho_{tW^-})$ with and without the cut $|\cos \theta_t| \leq 0.2$, for several values of the number of reconstructed $t \bar t$ events in that region. The statistical uncertainties are obtained by performing 1000 pseudo-experiments in which (i) a random set of events is selected; (ii) the matrices $\rho$ and $\rho^{T_B}$ are computed from this set; (iii) the procedure is repeated 1000 times. The negativity is obtained directly using Eq.~(\ref{ec:Nrho}).

\begin{table}[t]
\begin{center}
\begin{tabular}{ccc}
events & $|\cos \theta_t| \leq 0.2$ & no cut \\
5000  & $0.321 \pm 0.023$ & $0.179 \pm 0.024$ \\
10000 & $0.313 \pm 0.015$ & $0.170 \pm 0.017$ \\
20000 & $0.310 \pm 0.010$ & $0.164 \pm 0.011$ \\
\end{tabular}
\caption{Expected statistical uncertainty on $N(\rho_{tW^-})$ (central value and standard deviation)  obtained performing pseudo-experiments, for $e_R^+ e_L^-$ collisions at 1 TeV with $\cos \theta \leq -0.9$.} 
\label{tab:stat}
\end{center}
\end{table}

Observing $N(\rho_{tW^-}) > N(\rho_{t \bar t})$ with a statistical uncertainty of five standard deviations needs around $2 \times 10^4$ reconstructed events with the cut $|\cos \theta_t| \leq 0.2$, and 70000 events without that requirement. (We neglect the statistical uncertainty on $N(\rho_{t \bar t})$, which can be measured using a much larger sample.) Since  $|\cos \theta_t| \leq 0.2$ reduces the cross section by a factor of 6, the latter case is then more favourable. Given the small cross section for $t \bar t$ production, having such number of events requires a luminosity of the order of 100 ab$^{-1}$, which is far above the current proposals for near-future $e^+ e^-$ colliders. An optimisation of the measurement region in $\theta_t$, $\theta$, $\phi$ will reduce the luminosity required, but a detailed study is out of the scope of this work.


\section{Summary}
\label{sec:7}

In this work we have used the formalism of post-decay density operators~\cite{Aguilar-Saavedra:2024fig} and newly-calculated polarised decay amplitudes to investigate post-decay $t \bar t$ entanglement in detail. Top pair production is the best suited process to study post-decay entanglement at the LHC and thus perform novel tests of quantum mechanics.

We have investigated the factors that allow to have a sizeable post-decay entanglement. It is precisely the relatively low spin analysing power of the $W$ boson that allows to have a large $t W^- \bar b$ (or equivalently $\bar t W^+ b$) entanglement. Moreover, as previously pointed out~\cite{Aguilar-Saavedra:2023hss} the chirality of the $tbW$ coupling and smallness of the bottom quark mass allow the entanglement to be maintained when the $\bar b$ spins are traced out, leading to $tW^-$ entanglement. 

We have provided predictions for $tW^-$ entanglement at the LHC. Extending Ref.~\cite{Aguilar-Saavedra:2023hss}, we have found that the entanglement is nearly the same for any direction of the $W^-$ boson momentum $\vec p$. This fact may prove an advantage from the experimental point of view, to overcome possible efficiency and reconstruction issues. Actual measurements must consider a relatively narrow phase space region for $\vec p$ but, as we have shown here, the orientation of this region may be chosen at convenience. Moreover, measurements on different non-overlapping cones may be statistically combined. 

Finally, we have given predictions for polarised $e^+ e^-$ collisions. The striking possibility to have a $tW^-$ entanglement larger than the $t \bar t$ one~\cite{Aguilar-Saavedra:2024fig} has been investigated. An experimental observation could in principle be possible, but it would require very large statistics, beyond current proposals for near-term future colliders.

\section*{Acknowledgements}
I thank J. A. Casas for many discussions and previous collaboration.
This work has been supported by the Spanish Research Agency (Agencia Estatal de Investigaci\'on) through projects PID2019-110058GB-C21, PID2022-142545NB-C21, and CEX2020-001007-S funded by MCIN/ AEI/10.13039/501100011033, and by Funda\c{c}{\~a}o para a Ci{\^e}ncia e a Tecnologia (FCT, Portugal) through the project CERN/FIS-PAR/0019/2021.


\appendix

\section{Comparison with Monte Carlo simulation}
\label{sec:a}

We compare here the semi-analytical predictions obtained with the formalism of post-decay density operators and direct Monte-Carlo calculations. This serves not only as a further check of the formalism, but also to test the influence of finite top and $W$ width effects. 
For better comparison we use $gg \to t \bar t$ with $m_{t \bar t} \leq 370$ GeV. The reason is that the diagonalisation of the $6 \times 6$ matrices is rather sensitive to small perturbations originating from statistics, and even with a Monte Carlo sample of 6.9 million $t \bar t$ pairs, the restriction of the top anti-quark decay phase space results in relatively small samples in some cases. The $t \bar t$ density operator is given by
\begin{equation}
\rho = \left( \! \begin{array}{cccc}
0.062 & 0 & 0 & 0 \\
0 & 0.438 & -0.402 & 0 \\
0 & -0.402 & 0.438 & 0 \\
0 & 0 & 0 & 0.062
\end{array} \! \right)
\end{equation}
We use the same kinematical regions in decay phase space studied in section~\ref{sec:5}. In the numerical simulation, the condition of $\vec p$ being parallel to any axis $\hat n$ is implemented by requiring $\cos (\vec p,\hat n) \geq 0.99$. Results are presented in table~\ref{tab:ent3}. In order to get a sense of the numerical uncertainties present in the simulation, we present in table~\ref{tab:nev} the number of events in each kinematical region. Notice, however, that the uncertainty in the matrix diagonalisation does not scale as $1/\sqrt{N}$, with $N$ the number of events. Overall, we find excellent agreement between the semi-analytical and numerical computations in all cases.

\begin{table}[htb]
\begin{center}
\begin{tabular}{llclc}
& Principal eigenvector & $\lambda_\text{max}$ \\
$\vec p \, \| \hat z$ & $0.844 \UZ - 0.537 \DU$ & 0.846 \\
                    & $0.835 \UZ - 0.549 \DU$ & 0.825 \\
$\vec p \, \| -\hat z$ & $0.538 \UD - 0.843 \DZ$ & 0.846 \\
                    & $0.551 \UD - 0.834 \DZ$ & 0.840 \\
$p_3 > 0$ & $0.676 \UZ - 0.737 \DU$ & 0.561 \\
          & $0.664 \UZ - 0.747 \DU$ & 0.560 \\
$p_3 < 0$ & $0.738 \UD - 0.675 \DZ$ & 0.561 \\
          & $0.746 \UD - 0.665 \DZ$ & 0.560 \\
$\vec p \,\|\hat x$ &  $0.611 [\DU - \UD] + 0.238 [\UU - \DD] + 0.264 [\DZ - \UZ]$ & 0.850\\
                     & $0.610 [\DU - \UD] + 0.243 [\UU - \DD] + 0.261 [\DZ - \UZ]$ & 0.831 \\
$\vec p \,\|-\hat x$ & $0.611 [\DU + \UD]- 0.238 [\UU + \DD] - 0.264 [\DZ + \UZ]$ & 0.844 \\
                     & $0.612 [\DU + \UD]- 0.230 [\UU + \DD] - 0.265 [\DZ + \UZ]$ & 0.844 \\
$p_1 > 0$ & $0.598 [\DU - \UD] + 0.077 [\UU - \DD] + 0.370 [\DZ - \UZ]$ & 0.561 \\
          & $0.600 [\DU - \UD] + 0.079 [\UU - \DD] + 0.365 [\DZ - \UZ]$ & 0.558 \\
$p_1 < 0$ & $0.598 [\DU + \UD] - 0.077 [\UU + \DD] - 0.370 [\DZ + \UZ]$ & 0.561 \\
          & $0.600 [\DU + \UD] - 0.079 [\UU + \DD] - 0.365 [\DZ + \UZ]$ & 0.559 \\
$\vec p \,\| \hat y$ & $0.611 [\DU - i \UD] + 0.238 [\DD - i \UU] - 0.263 [\UZ - i \DZ]$ & 0.850 \\
                     & $0.614 [\DU - i \UD] + 0.232 [\DD - i \UU] - 0.262 [\UZ - i \DZ]$ & 0.837 \\
$\vec p \, \| - \hat y$ & $0.611 [\DU + i \UD] + 0.238 [\DD + i \UU] - 0.263 [\UZ + i \DZ]$ & 0.850 \\
                        & $0.610 [\DU + i \UD] + 0.244 [\DD + i \UU] - 0.258 [\UZ + i \DZ]$ & 0.831 \\
$p_2 > 0$ & $0.597 [\DU - i \UD] + 0.077 [\DD - i \UU] - 0.370 [\UZ - i \DZ]$ & 0.561 \\
          & $0.601 [\DU - i \UD] + 0.079 [\DD - i \UU] - 0.365 [\UZ - i \DZ]$ & 0.557 \\
$p_2 < 0$ & $0.597 [\DU + i \UD] + 0.077 [\DD + i \UU] - 0.370 [\UZ + i \DZ]$ & 0.561 \\
          & $0.601 [\DU + i \UD] + 0.078 [\DD + i \UU] - 0.364 [\UZ + i \DZ]$ & 0.558
\end{tabular}
\caption{Principal eigenvector and largest eigenvalue of the $tW^-$ spin density operator, for selected kinematical regions. In each case, the first row corresponds to the semi-analytical calculation and the second row to the Monte Carlo simulation. We write in compact notation the $1/2$ and $1$ spin eigenvalues of $t$ and $W^-$ as `$+$', and the $-1/2$, $-1$ eigenvalues as `$-$'.} 
\label{tab:ent3}
\end{center}
\end{table}

\begin{table}[htb]
\begin{center}
\begin{tabular}{lclclc}
$\vec p \, \| \hat z$ & 36772 & $\vec p \,\|\hat x$ & 33341 & $\vec p \,\| \hat y$ & 33371 \\
$\vec p \, \| -\hat z$ & 36865 & $\vec p \,\|-\hat x$ & 33560 & $\vec p \, \| - \hat y$ & 33316 \\
$p_3 > 0$ & 3435992 & $p_1 > 0$ & 3434559 & $p_2 > 0$ & 3436644 \\
$p_3 < 0$ & 3431355 & $p_1 < 0$ & 3432788 & $p_2 < 0$ & 3430703
\end{tabular}
\caption{Number of events in the numerical simulation for each of the kinematical regions considered.}
\label{tab:nev}
\end{center}
\end{table}

\end{document}